# Strain Sensitivity in the Nitrogen 1*s* NEXAFS Spectra of Gallium Nitride

Andrew Ritchie[1], Shaylin Eger[1], Chelsey Wright[2], Daniel Chelladurai,[1,#] Cuyler Borrowman[1], Weine Olovsson[3], Martin Magnuson[3], Jai Verma[4], Debdeep Jena[4], Huili Grace Xing[4], Christian Dubuc[5], Stephen Urquhart[1,]*

1. *Department of Chemistry, University of Saskatchewan, Saskatoon, SK, Canada*

2. *Canadian Light Source, Saskatoon, SK, Canada*

3. *Department of Physics, Chemistry and Biology (IFM), Linköping University, SE-581 83 Linköping, Sweden*

4. *Department of Electrical Engineering, University of Notre Dame, Notre Dame, IN, USA*

5. *Osemi Canada Inc., Sherbrooke, Quebec, Canada*

**Abstract**

The Nitrogen 1*s* near edge X-ray absorption fine structure (NEXAFS) of gallium nitride (GaN) shows a strong natural linear dichroism that arises from its anisotropic wurtzite structure. An additional spectroscopic variation arises from lattice strain in epitaxially grown GaN thin films. This variation is directly proportional to the degree of strain for some spectroscopic features. This strain variation is interpreted with the aid of density functional theory calculations.

\* Corresponding Author: stephen.urquhart@usask.ca

# Current address, Nanotechnology Engineering Program, University of Waterloo, Waterloo, ON

## 1. Introduction

Gallium nitride (GaN) has potential use for semiconductor and optoelectronic devices in high power and high temperature applications.[1-3] Commercial GaN device applications are under pressure to reduce costs and maintain compatibility with the enormous infrastructure based on silicon. A significant complication for the commercial fabrication of GaN devices on silicon is strain. The lattice mismatch between GaN and silicon is large, causing significant strain in GaN films grown on silicon (16.9% for GaN films grown on Si(111) substrates, relative to 4% for GaN films grown on SiC).[4] This strain affects the mechanical stability of the GaN film, leading to cracking and physical distortion.[5] To reduce the negative consequences of strain, complex buffer or nucleation layer techniques are used. Semiconductor strain can also be used for band structure engineering in GaN-based materials, in order to optimize their light emitting properties.[2,6]

Spatially resolved characterization of strain in GaN semiconductor devices has relevance for informing the development of low-stress GaN deposition methods, for studying the effects of processing on strain in GaN devices, and for optimizing GaN devices that incorporate engineered strain to obtain desired optoelectronic properties. To accurately map strain at the current and future size scales of semiconductor devices, down to the size scale of individual gates, the spatial resolution of strain mapping must be increased.

This work aims to characterize the strain sensitivity of Near Edge X-ray Absorption Fine Structure (NEXAFS) spectroscopy, as a foundational step towards the development of a high spatial resolution strain metrology based on X-ray microscopy. This potential strain metrology technique would combine the high spatial resolution of X-ray microscopy with the strain sensitivity of NEXAFS spectroscopy. For example, aberration corrected X-ray Photoemission Electron Microscopy (X-PEEM) has the potential for using NEXAFS image contrast with an ultimate lateral spatial resolution below 10 nm.[7] This is significantly below the spatial resolution of most comparable techniques, while offering rapid, full-field, and near-surface sensitive metrology. In comparison, Raman microscopy is effective for spatially resolved characterization of strain in GaN semiconductors,[8,9] allowing for rapid strain characterization with ~ 0.25 micron spatial resolution.[10,11] However, the sampling depth of Raman spectroscopy restricts its use to films thicker than the Raman sampling depth when the substrate can interfere. In our work, Raman scattering from AlN and sapphire in the substrate interfered with the signal from the GaN thin film (10 – 46 nm thick). In contrast, the depth sensitivity of NEXAFS is on the scale of a few nanometers, providing a surface sensitive probe.[12] Micro-XRD has a spatial resolution similar to Raman (~0.25 micron).[13] TEM is capable of imaging lattice strain with atomic resolution, but requires extensive and destructive sample preparation work.[14] Tip enhanced Raman spectroscopy can be used for nanometer scale strain mapping, as demonstrated by studies of SiGe quantum dots.[15] However, this technique is not conducive to rapid data acquisition.

Towards the goal of an X-ray microscopy based strain metrology, the strain sensitivity of NEXAFS spectroscopy must first be determined. This is the goal of this work. Previously, we studied the sensitivity of natural linear dichroism[16] in the Silicon 1*s* NEXAFS spectroscopy to strain in SiGe alloys. We found that the magnitude of the *linear dichroism difference* (the difference between NEXAFS spectra recorded with the X-ray linear polarization directed in the sample plane and ~normal to the surface) was linearly proportional to the degree of strain.[17]

A similar strain effect is sought for GaN semiconductors. As unstrained Si and SiGe alloys have cubic structures, the Silicon 1$s$ NEXAFS spectrum will only show a linear dichroism difference when strain is present. In contrast, the anisotropic wurtzite structure of GaN has a strong intrinsic linear dichroism[3,18,19] in the absence of strain. As strain is manifest as an expansion or compression of the *ab* plane with a corresponding truncation or elongation of the *c* axis, strain should then modify the intrinsic linear dichroism of GaN.

This paper explores the sensitivity of Nitrogen 1$s$ NEXAFS spectroscopy to the degree of strain in GaN thin films. NEXAFS provides an element-specific probe of the unoccupied electronic structure of a material. Several different core edges are possible for GaN: Gallium 1$s$ (~ 10370 eV), Gallium 2$p$ (~ 1140 eV) and Nitrogen 1$s$ (~ 410 eV).[3] Features in the Nitrogen 1$s$ NEXAFS spectra are narrower and better resolved than in the Gallium 1$s$ and 2$p$ spectra[3] on account of the longer lifetime of the Nitrogen 1$s$ core excited state. On this basis, the Nitrogen 1$s$ edge spectra will be more sensitive to strain.

## 2. Experimental

### 2.1 Sample Preparation

GaN thin films of varying thickness (10 nm to 46 nm) were grown by molecular beam epitaxy on AlN / sapphire substrates (~1.1 μm AlN on sapphire) at the University of Notre Dame.

GaN samples were grown with a Veeco Gen 930 MBE system with RF plasma source for Nitrogen sources and effusion cells for metal sources. AlN on sapphire substrates were procured commercially. The templates were In mounted on Si wafers. After being loaded into the MBE system, substrates were baked at 200 $^\circ$C to desorb water vapor, oxygen, $CO_2$, etc. from their surfaces. Prior to growth, the substrates were baked again at 450 $^\circ$C to desorb any adsorbed gases. A substrate temperature of 730 °C and RF plasma power of 220 W, corresponding to a growth rate of 150 nm/hr, was employed. Initially, a 50 nm AlN layer was grown at stoichiometric conditions (Al/N~1) to bury the growth interface, which can be defective. GaN layers of varying thicknesses were then grown in a metal rich regime (Ga/N > 1) to obtain smooth films. Streaky reflection high-energy electron diffraction (RHEED) patterns corroborated the growth of smooth films in this regime. As Raman spectroscopy was ineffective as a strain measurement due to interferences from the AlN/sapphire substrate, strain was directly measured using reciprocal X-ray diffraction (XRD) mapping. A Panalytical XRD system equipped with an X-ray source emitting at 45 keV was utilized for XRD scans. The ω-2θ triple axis reciprocal scan was performed along the (1 0 5), off–axis plane to measure the strain in GaN films with respect to AlN.

A summary of sample information is provided in **Table 1**. As shown by **Equation 1**, 100 % strain refers to the GaN film adopting the AlN in-plane lattice structure, while 0 % strain refers to the fully relaxed GaN structure.

(Equation 1) $\quad\text{Strain} = \frac{(a_{GaN}-a_m)}{(a_{GaN}-a_{AlN})} x 100\%$

where $a_{GaN}$ and $a_{AlN}$ are the $a$-lattice constants of relaxed GaN and AlN (3.190 Å and 3.110, respectively),[20] and $a_m$ is the measured $a$-lattice constant of the strained GaN films.

## 2.2 NEXAFS Spectroscopy

Nitrogen 1$s$ NEXAFS spectra of GaN thin films were obtained in the X-PEEM endstation on the spectromicroscopy (SM) beamline at the Canadian Light Source (CLS),[21] without spatial resolution. Preliminary data were also acquired on the Hermon beamline at the Synchrotron Radiation Center (SRC) and on the SGM beamline at the CLS.

NEXAFS spectra were acquired by measuring the averaged X-PEEM image intensity as a function of photon energy. The X-PEEM image comes from electrons emitted from the sample surface that are accelerated and magnified onto an image intensifier screen. This provides the same signal as total electron yield (TEY) detection except that the signal is measured through the X-PEEM imaging optics rather than the sample drain current. These X-PEEM images can be used to extract spatially resolved spectroscopy, but in these experiments on featureless samples, the entire X-PEEM field of view was averaged. The sampling depth of this detection method is ~5 nm,[12] which is less than the thickness of the thinnest GaN film.

NEXAFS spectra were normalized by the total yield spectrum of the Au-plated Ni mesh ($I_\circ$ measurement) in order to remove the monochromator function. NEXAFS were background subtracted and normalized in the continuum (440 – 445 eV) to the atomic absorption cross-section of Nitrogen,[22] following the method outlined by Hitchcock and Mancini.[23] This normalization is illustrated in the accompanying data to this paper. The energy resolution at the Nitrogen 1$s$ edge for these experiments is approximately 0.2 eV.

The experimental geometry used in these experiments is presented in **Figure 1**. The sample was oriented at a glancing angle to the X-ray beam (16° from the surface / *ab* plane), with the surface normal of the sample in horizontal. An Apple II Elliptically Polarizing Undulator (EPU) was used to control the linear polarization of the X-ray beam. The EPU was used to change the polarization from parallel (E parallel to the sample surface, the *ab* plane) to perpendicular (E ~ perpendicular to the sample surface), as well as three inclinations between (22.5° steps).

## 2.3 Computational Studies

First-principles DFT[24],[25] calculations have been used to further understand the strain dependence in the NEXAFS spectra of GaN. The Nitrogen 1*s*-edge NEXAFS spectra of strained and unstrained GaN was performed utilizing the all-electron full potential linear augmented plane wave (FPLAPW) WIEN2k program package.[26] Supercell techniques are computationally efficient, and provide generally good agreement with experiment.[27] Lattice constants were taken from experimental XRD data. For these WIEN2k calculations, the augmented plane wave + local orbital (APW+lo) basis set was used. The exchange-correlation functional used was the general gradient approximation (GGA), according to Perdew *et al*.[28] The core-hole approximation, wherein a full electron is removed from the 1*s* orbital at a single Nitrogen site and an extra electron is added into the valence band to keep charge neutrality of the supercell, was used. Self-consistent calculations were performed and Fermi's golden rule was used to compute the NEXAFS spectra. Spectra were produced using an 8 x 8 x 4 *k*-mesh with 35 inequivalent *k*-points and a 108-atom supercell in order to avoid unphysical interactions between core-ionized sites due to periodic boundary conditions. For comparison with experiment, a Gaussian broadening was applied to the theoretical spectra and theoretical spectra were aligned with the corresponding features in experiment by a rigid shift of 398 eV. Theoretical spectra were normalized to match their maximum intensities to those of their corresponding experimental spectra.

## 3. Results and Discussion

**Figure 2** presents the normalized Nitrogen 1*s* NEXAFS of a fully strained, 10 nm thick GaN sample as a function of the orientation of the X-ray linear polarization relative to the sample. The linear dichroism in the Nitrogen 1*s* NEXAFS spectra of GaN is apparent, particularly for the three low energy peaks, labeled 1-3. The intensities of peak 1 and peak 3 are the least intense when the X-ray polarization is parallel to the *ab* plane and most intense when perpendicular. The polarization dependence of peak 2 is less consistent. The change in the energy of peak 2 with the polarization angle indicates that at least two transitions, with differing polarization dependencies, underlie this feature. Higher energy features in the spectra are broad and also vary with polarization angle. As broad features are less sensitive to strain, they will not be considered further.

As strain is a form of structural anisotropy, the polarization dependence is expected to change with strain. **Figure 3** presents the perpendicular (3a, top) and parallel (3b, bottom) –polarized spectra recorded for a set of GaN samples with different degrees of strain. The effect of strain is obvious in peaks 1 and 3 in the perpendicular polarized spectra, increasing in intensity with strain. Peak 2 shows a less consistent strain dependence, as expected as this feature contains multiple overlapping contributions. The variation of the parallel polarization spectra with strain is smaller.

The observation of a variation in the perpendicular polarized N 1*s* spectra with strain is remarkable. This result also indicates that strain in GaN can be measured without taking a linear dichroism difference (the differences between perpendicular and parallel polarized spectra), as required in our earlier strained SiGe study.[17]

The variation in the experimental peak intensities with strain for the perpendicularly polarized spectra is presented in **Figure 4.** Data is also shown for theoretical calculations that follow. The areas underlying peak 1 increase monotonically with the degree of GaN strain. The relationship for peak 3 is less clear, probably due to the presence of other underlying features at this energy.

**Figure 5** presents a direct comparison between the theoretical and experimental Nitrogen 1*s* NEXAFS spectra for a 90 % strained GaN sample, for perpendicular and parallel polarizations. The calculated spectra have somewhat different shapes than the experimental spectra, in part due to the absence of a photoionization background in the calculated spectra. However, the general shape of the lower energy band is well reproduced. The calculated spectra also show additional fine features (such as at ~ 404 eV in perpendicular spectra) that are not observed in the broader experimental spectra.

**Figure 6** presents a comparison of calculated Nitrogen 1*s* NEXAFS spectra of GaN for three different levels of strain. The variation in the intensity of the first peak with the degree of strain is reproduced in both the parallel and perpendicular polarized spectra. Theory, with its ability to resolve features inaccessible to experiment, shows the origin of the complex behavior of peak 2 shown in **Figure 2**. Two features contribute to the energy shift of this band with polarization change. As there are fewer apparent spectroscopic contributions to peak 1, this is the best feature for spectroscopic characterization of strain in GaN.

The origin of the strain dependence of the first peak in the N 1*s* NEXAFS spectra can be related to a change in the GaN structure. At higher strain (as GaN adopts the smaller AlN *ab* lattice constant), the *ab* plane of GaN is compressed and the *c* axis expands; this is shown in **Table 1**. Through the dipole selection rule, the perpendicular polarized Nitrogen 1*s* spectra will sample 2*p* states that are perpendicular to the *ab* plane (parallel to *c*). As strain increases, the single Ga-N bond along the *c* axis will increase, and the pyramidal base formed by the three Ga-N bonds in the *ab* plane will slightly compress. With the increase in the Ga-N bond along the *c*-axis, a transition with Nitrogen 1*s* → π* character directed along the *c*-axis should become more intense, as observed. An illustration of the nitrogen geometry in wurtzite GaN that is modified by this strain is found in Amidani *et al*.[29]

## 4. Conclusions

In this study, the strain dependence in the polarization dependent Nitrogen 1*s* NEXAFS spectra of GaN thin films was examined. A weak strain dependent effect is observed, superimposed on the natural linear dichroism in these spectra. The intensity of the first feature in the perpendicularly polarized Nitrogen 1*s* spectrum of GaN scales linearly with the degree of polarization, while others show a more complex variation. Theoretical DFT calculations reproduce these strain dependent trends.

This is the first known observation of strain dependence in the NEXAFS spectra of GaN. This strain dependence could provide a contrast mechanism for a high spatial resolution strain metrology, based on X-ray microscopy.

Applied Surface Science **316**, 232 (2014)

**Acknowledgements**

Research described in this paper was performed at the Canadian Light Source, which is supported by NSERC, NRC (Canada), CIHR, Province of Saskatchewan, WED, and the University of Saskatchewan. Research was also performed at Synchrotron Radiation Center which was primarily funded by the University of Wisconsin-Madison with supplemental support from facility Users and the University of Wisconsin-Milwaukee. SGU wishes to thank the Synchrotron Radiation Center staff for their many years of support and scientific leadership. SGU acknowledges financial support from the NSERC Canada under a Strategic Grant program. W.O. and M.M. acknowledge funding from the Linköping Linnaeus Initiative for Novel Functional Materials (LiLi-NFM) supported by the Swedish Research Council (VR) and the VR grant number 621-2011-4426. The calculations were carried out at the National Supercomputer Centre (NSC) at Linköping University supported by the Swedish National Infrastructure for Computing (SNIC). HGX acknowledges funding from the National Science Foundation (NSF) grant number DMR-0907583.

**Table 1**. GaN sample details: lattice parameters, film thicknesses and lattice strain (defined relative to AlN, see Equation 1).

| Strain / % | GaN Sample Thickness / nm | $a$ / Å | $c$ / Å |
|---|---|---|---|
| 71 | 46 | 3.134 | 5.227 |
| 90 | 14.8 | 3.120 | 5.242 |
| 100 | 10 | 3.112 | 5.242 |

**References**

[1]  H. Morkoç, S. Strite, G.B. Gao, M.E. Lin, B. Sverdlov, M. Burns, Journal of Applied Physics 76 (1994) 1363.
[2]  H. Jia, L. Guo, W. Wang, H. Chen, Advanced Materials 21 (2009) 4641.
[3]  M. Magnuson, M. Mattesini, C. Höglund, J. Birch, L. Hultman, Physical Review B 81 (2010) 085125.
[4]  C. Kisielowski, J. Krüger, S. Ruvimov, T. Suski, J.W. Ager, III, E. Jones, Z. Liliental-Weber, M. Rubin, E.R. Weber, M.D. Bremser, R.F. Davis, Physical Review B 54 (1996) 17745.
[5]  E. Feltin, B. Beaumont, M. Laügt, P. de Mierry, P. Vennéguès, H. Lahrèche, M. Leroux, P. Gibart, Applied Physics Letters 79 (2001) 3230.
[6]  S.L. Wang, B.C. Yeh, H.M. Wu, L.H. Peng, C.M. Lai, T.S. Ko, T.C. Lu, S.C. Wang, A.H. Kung, physica status solidi (c) 5 (2008) 1780.
[7]  T. Schmidt, U. Groh, R.F. Fink, E. Umbach, O. Schaff, W. Engel, B. Richter, H. Kuhlenbeck, R. Schlögl, H.J. Freund, A.M. Bradshaw, D. Preikszas, P. Hartel, R. Spehr, H. Rose, G. Lilienkamp, E. Bauer, G. Benner, Surface Review and Letters 9 (2002) 223.
[8]  G.H. Loechelt, N.G. Cave, Journal of Applied Physics 86 (1999) 6164.
[9]  M. Katsikini, J. Arvanitidis, D. Christofilos, S. Ves, G.P. Dimitrakopulos, G. Tsiakatouras, K. Tsagaraki, A. Georgakilas, physica status solidi (a) 209 (2012) 1085.
[10]  M. Kuball, M. Benyoucef, B. Beaumont, P. Gibart, Journal of Applied Physics 90 (2001) 3656.
[11]  N. Hayazawa, M. Motohashi, Y. Saito, S. Kawata, Applied Physics Letters 86 (2005) 263114.
[12]  J. Stohr, NEXAFS Spectroscopy, Springer-Verlag, 1992.
[13]  I. De Wolf, V. Senez, R. Balboni, A. Armigliato, S. Frabboni, A. Cedola, S. Lagomarsino, Microelectronic Engineering 70 (2003) 425.
[14]  C.-P. Liu, J.M. Gibson, D.G. Cahill, T.I. Kamins, D.P. Basile, R.S. Williams, Physical Review Letters 84 (2000) 1958.
[15]  Y. Ogawa, T. Toizumi, F. Minami, A. Baranov, Physical Review B 83 (2011).
[16]  S.E. Braslavsky, Pure Applied Chemistry 79 (2007) 293.
[17]  W. Cao, M. Masnadi, S. Eger, M. Martinson, Q.F. Xiao, Y.F. Hu, J.M. Baribeau, J.C. Woicik, A.P. Hitchcock, S.G. Urquhart, Applied Surface Science 265 (2013) 358.
[18]  K. Lawniczak-Jablonska, T. Suski, Z. Liliental-Weber, E.M. Gullikson, J.H. Underwood, R.C.C. Perera, T.J. Drummond, Applied Physics Letters 70 (1997) 2711.
[19]  M. Katsikini, E.C. Paloura, T.D. Moustakas, Journal of Applied Physics 83 (1998) 1437.
[20]  H. Schulz, K.H. Thiemann, Solid State Communications 23 (1977) 815.
[21]  K.V. Kaznatcheev, C. Karunakaran, U.D. Lanke, S.G. Urquhart, M. Obst, A.P. Hitchcock, Nuclear Instruments and Methods in Physics Research Section A: Accelerators, Spectrometers, Detectors and Associated Equipment 582 (2007) 96.
[22]  B.L. Henke, E.M. Gullikson, J.C. Davis, Atomic Data and Nuclear Data Tables 54 (1993) 181.
[23]  A.P. Hitchcock, D.C. Mancini, Journal of Electron Spectroscopy and Related Phenomena 67 (1994) vii.
[24]  P. Hohenberg, W. Kohn, Physical Review 136 (1964) B864.
[25]  W. Kohn, L.J. Sham, Physical Review 140 (1965) A1133.

[26]   K.S. P. Blaha, G Madsen, D. Kvasnicka and J Luitz, WIEN2k An Augmented Plane Wave Plus Local Orbitals Program for Calculating Crystal Properties, Vienna, Austria, 2001.
[27]   T. Mizoguchi, W. Olovsson, H. Ikeno, I. Tanaka, Micron 41 (2010) 695.
[28]   J.P. Perdew, K. Burke, M. Ernzerhof, Physical Review Letters 77 (1996) 3865.
[29]   R. Amritendu, M. Somdutta, S. Surajit, A. Sushil, P. Rajendra, G. Rajeev, G. Ashish, Journal of Physics: Condensed Matter 24 (2012) 435501.

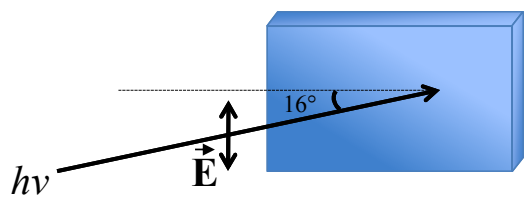

**Parallel:** *E* parallel to sample surface (*ab* plane)

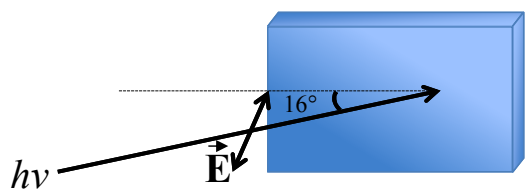

**Perpendicular:** *E* ~ perpendicular to sample surface (*c* axis)

**Figure 1:** Experimental geometry for the angle dependent NEXAFS experiments. **Parallel (top)**: *E* parallel to the sample surface (*ab* plane). **~Perpendicular (bottom)**: *E* ~perpendicular to the sample surface (*c* axis). **Not shown**: *E* in increments of 22.5° between the parallel and perpendicular geometries.

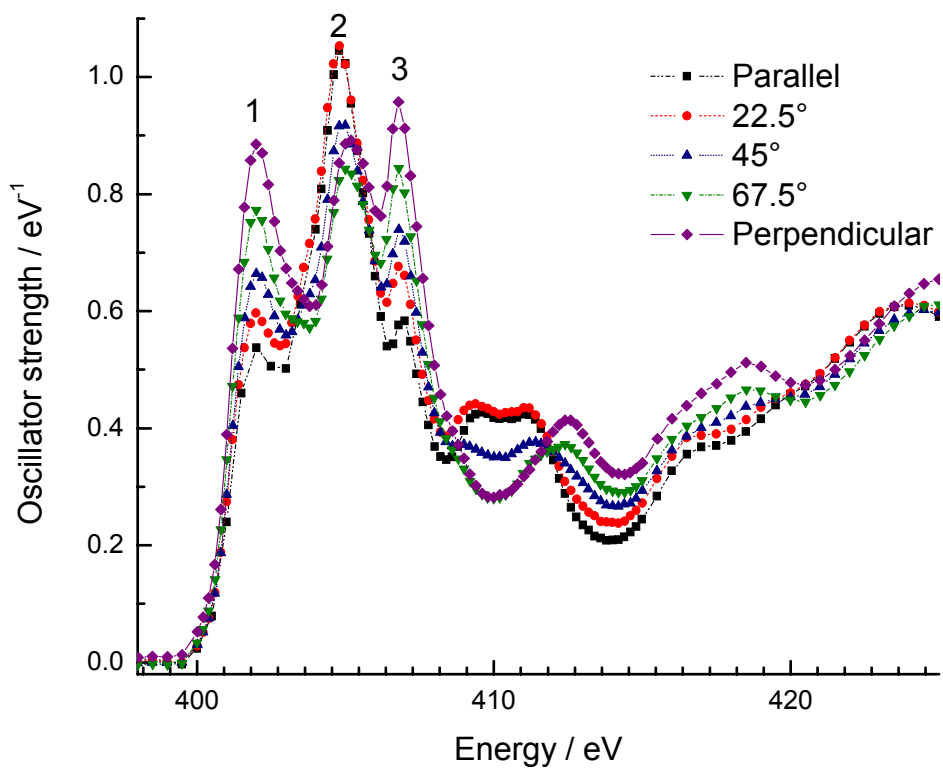

**Figure 2:** Nitrogen 1*s* NEXAFS spectra of a 100% strained, 10 nm thick GaN film grown on AlN. Spectra are presented for five polarization angles, indicated relative to the sample surface.

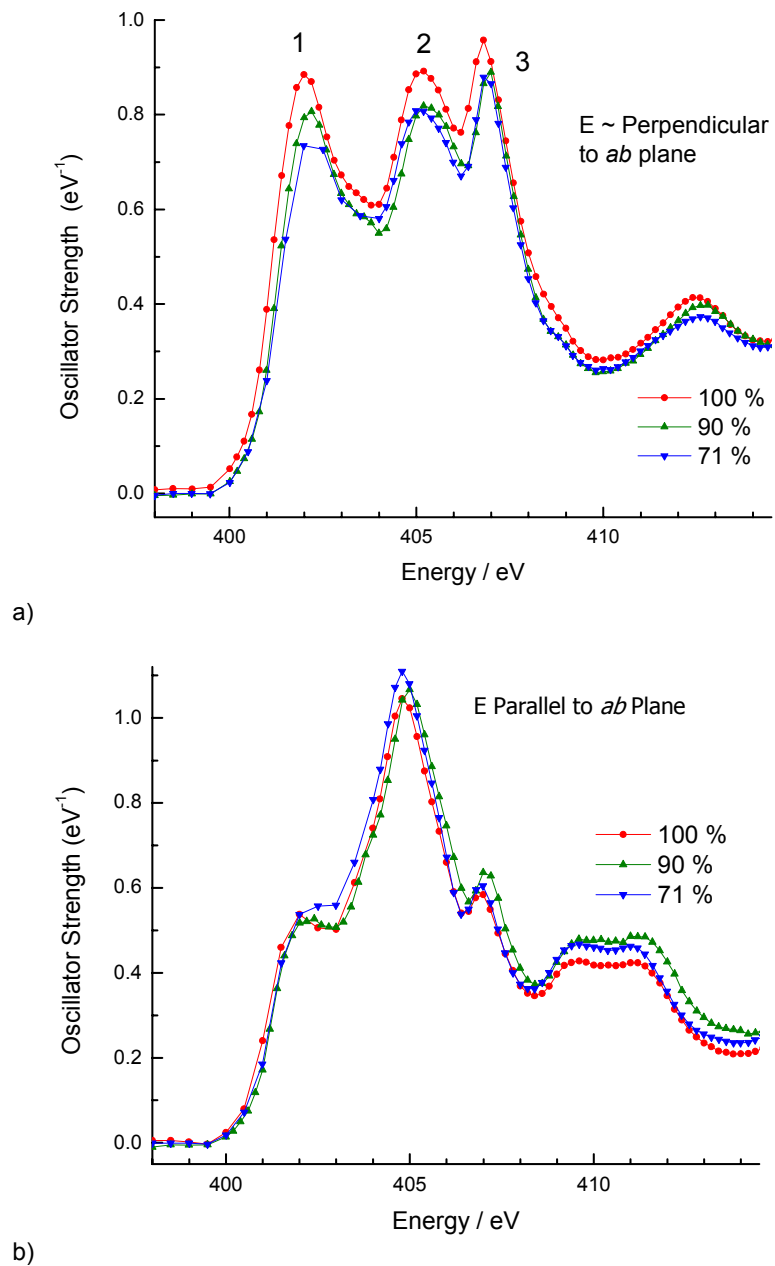

**Figure 3**: Perpendicular and parallel-polarized Nitrogen 1*s* NEXAFS spectra for a series of strained GaN samples. See Table 1 for sample details.

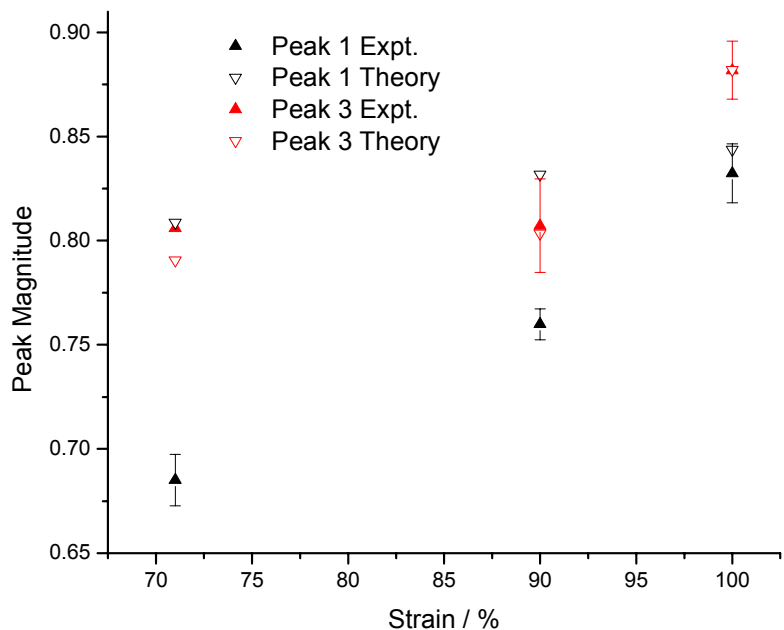

**Figure 4**: Correlation between the experimental and theoretical peak areas for Peaks 1 and 3 versus the degree of strain for a series of GaN thin films. The peak areas are extracted from the experimental and theoretical Nitrogen 1*s* spectra recorded with perpendicular polarization (E ~perpendicular to the *ab* plane). The areas are integrated over a range of maxima ± 0.5 eV. Theory values were scaled linearly for comparison to experiment. Experimental values are averages of three measured values and uncertainties shown are standard deviations.

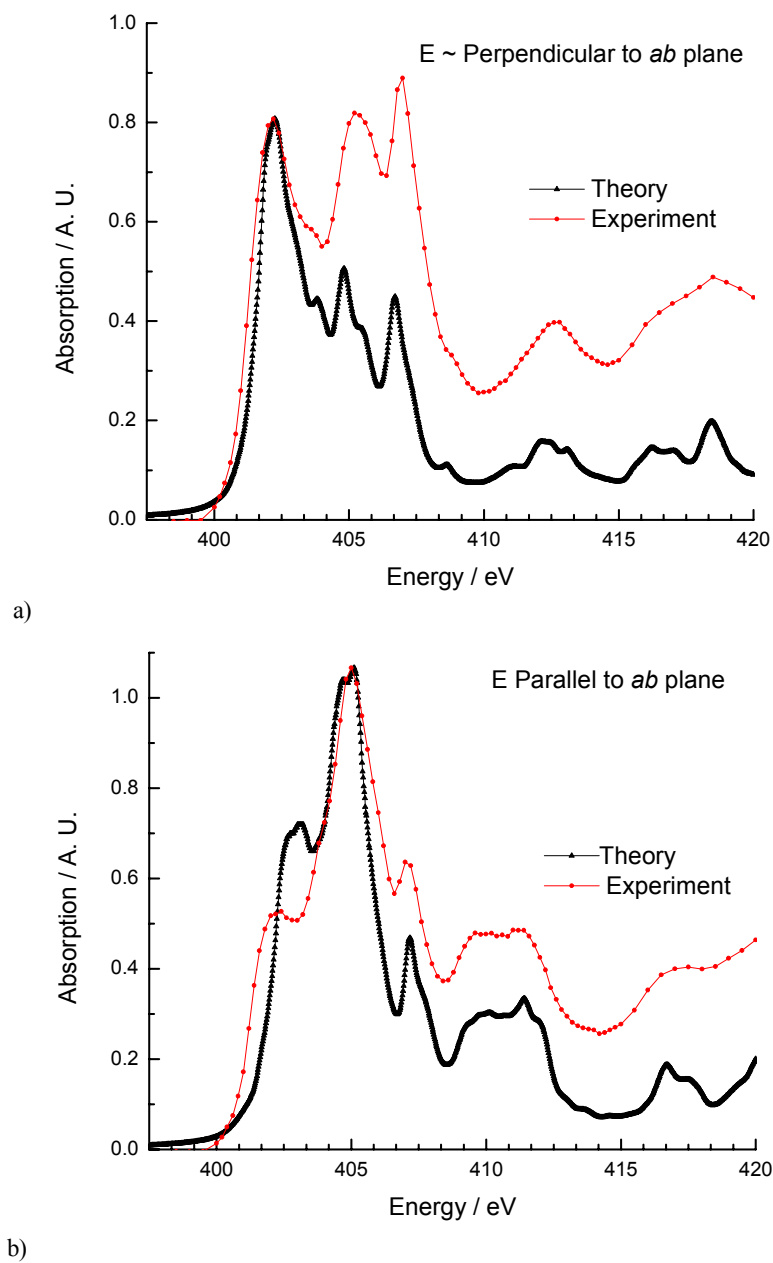

a)

b)

**Figure 5**: Comparison of calculated and experimental Nitrogen 1*s* NEXAFS spectra for the 90 % strained GaN sample (14.8 nm GaN on AlN), presented for perpendicular (*E* ~perpendicular to *ab*) and parallel polarizations (*E* parallel to *ab*). Theory values have been shifted by +398 eV for comparison on a common energy scale.]

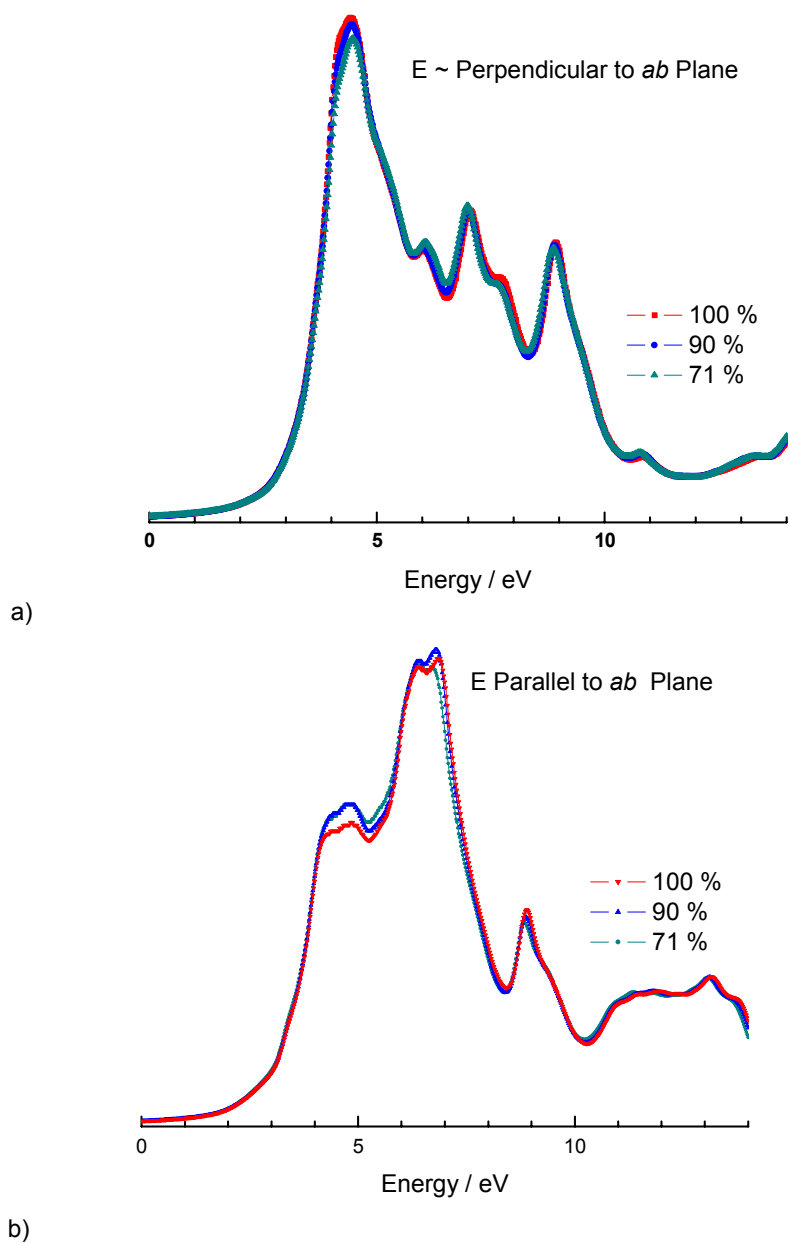

**Figure 6**: Simulated Nitrogen 1*s* NEXAFS spectra for several strained GaN models from DFT calculations, presented for perpendicular (*E* ~perpendicular to *ab*) and parallel polarizations (*E* parallel to *ab*).